\documentclass[aps,prl,superscriptaddress,twocolumn,nofootinbib]{revtex4-1}
\usepackage{chay,graphicx,amsmath}

\newcommand{\euv}{\epsilon_{\mathrm{UV}}}
\newcommand{\eir}{\epsilon_{\mathrm{IR}}}

\begin{document}

\title{Factorization theorem for high-energy scattering near the endpoint}

\def\KU{Department of Physics, Korea University, Seoul 136-713, Korea} 

\def\Seoultech{Institute of Convergence Fundamental Studies \& School of Liberal Arts, Seoul National University of Science and 
Technology, Seoul 139-743, Korea}
\author{Junegone Chay}
\email[E-mail:]{chay@korea.ac.kr}
\affiliation{\KU}†
\author{Chul Kim}
\email[E-mail:]{chul@seoultech.ac.kr}
\affiliation{\Seoultech\vspace{0.5cm}} 

\begin{abstract} \vspace{0.1cm}\baselineskip 3.0 ex 
A consistent factorization theorem is presented in the framework of effective field theories. Conventional factorization suffers 
from infrared divergences in the soft and collinear parts. We present a factorization theorem in which 
the infrared divergences appear only in the parton distribution functions by carefully  
reorganizing collinear and soft parts.  The central idea is extracting the soft contributions from the collinear part 
to avoid double counting. Combining it with the original soft part, an infrared-finite kernel is obtained. 
This factorization procedure can be applied to various high-energy scattering processes.
  
\end{abstract}

\maketitle

\baselineskip 3.0 ex 

Theoretical predictions on high-energy scattering processes are based on the factorization theorem, 
in which the cross section is factorized into hard, collinear and soft parts. 
The hard part consists of the partonic cross sections with hard momenta. The
collinear part describes the effects on the energetic particles. For hadron colliders, 
the collinear part involves the parton distribution functions (PDF). 
The soft part describes the soft gluon exchange between different collinear parts. 
These three parts are decoupled and the collinear and soft parts are defined
as the matrix elements of gauge-invariant operators.

This conventional factorization scheme, as it stands,  is plagued by infrared (IR) divergences in the collinear and soft parts. The 
hard part, by definition, depends on hard momenta and it is IR finite. There may exist ultraviolet (UV) divergences, but they can be 
removed by counterterms. The collinear and soft parts unavoidably encounter IR divergences in massless gauge theories
like QCD whatever the regulators are. If the dimensional regularization is employed to regulate UV and IR
divergences, the divergences are expressed in terms of the poles in $\epsilon_{\mathrm{UV}}$  ($\eir$) for UV (IR) divergences.
These UV and IR poles appear separately and there is no mixing between them. If the UV divergence is regulated
by the dimensional regularization, but the IR divergence is regulated by the offshellness of external particles, 
there are UV poles, and also logarithms of the offshellness for IR divergences.

Our idea of factorization goes further than the operator definition and its radiative corrections of each part in order to isolate IR divergences. 
The radiative corrections are reorganized
such that the IR divergences reside only in the PDFs, as in full QCD, and the remaining quantities are IR safe. Furthermore the method of
the reorganization is so simple and universal that it can be applied to a variety of scattering
processes. The goal is to find an IR-finite kernel to be convoluted with the hard function and the PDF  in the factorized 
expression of the cross section.
The evolution of the kernel can be consistently described using the renormalization group equation without devising a method to
handle IR divergences.  
We will delineate our method of factorization, and explain 
how it describes the factorization theorems in deep-inelastic scattering (DIS), and Drell-Yan (DY) processes near the endpoint 
or with transverse momentum.

We can schematically write the factorized scattering cross sections near the endpoint as
\begin{equation} \label{nfac}
d\sigma =\left\{ \begin{array}{l}
H(Q,\mu) \otimes S(E,\mu) \otimes f_1 (E,\mu) \otimes f_2 (E,\mu), \\
H(Q,\mu) \otimes J(Q \sqrt{1-z},\mu)\otimes S(E,\mu) \otimes f(E,\mu), 
\end{array}
\right.
\end{equation}
in DY and DIS processes respectively. $H$ is the hard function, and $S$ is the soft function, which is defined in terms of
 soft Wilson lines~\cite{Korchemsky:1993uz}. $f_1$, $f_2$ and $f$ 
are the initial collinear functions defined  in terms of the quark bilinear operators at the intermediate scale 
$E\sim Q (1-z)$. $J$ is the jet
function in DIS, which describes the collinear final-state particles where $z$ is the Bjorken variable.
Here $\otimes$ means an appropriate convolution, and $Q$ is the large scale in the scattering. 
The factorized form in Eq.~(\ref{nfac}) is well established in QCD~\cite{Sterman:1986aj,Catani:1989ne}.

Since the 
processes involve multiple scales, it is convenient and transparent to understand the physics by employing an effective theory, which 
is soft-collinear effective theory (SCET)~\cite{Bauer:2000yr,Bauer:2001yt}. The conventional factorization theorem, 
Eq.~(\ref{nfac}), can be rephrased exactly by noting that 
SCET factorizes the soft and collinear interactions from the outset~\cite{Chay:2005rz,Becher:2007ty}. The soft part is expressed in terms of the soft Wilson lines as
\begin{equation}\label{sdy}
S_{\mathrm{DY}} (E,\mu) = \frac{1}{N_c} \langle 0| \mathrm{tr} Y_n^{\dagger} 
Y_{\bar{n}} \delta (1-z + 2i\partial_0/Q) Y_{\bar{n}}^{\dagger} 
Y_n |0\rangle
\end{equation}
for DY process near the endpoint $z=Q^2/\hat{s} \sim 1$ where $\hat{s}$ is the partonic center-of-mass energy squared. And
\begin{equation} \label{sdis}
S_{\mathrm{DIS}} (E,\mu) = \frac{1}{N_c} \langle 0| \mathrm{tr} Y_n^{\dagger} \tilde{Y}_{\bar{n}} \delta (1-z + i\overline{n}\cdot 
\partial/Q) \tilde{Y}_{\bar{n}}^{\dagger} 
Y_n |0\rangle
\end{equation}
for DIS near the endpoint $z=Q/\overline{n}\cdot p \sim 1$, where $z$ is the partonic Bjorken variable. The typical soft scale is given by
$E\sim Q(1-z)$.
Here $n$ and $\overline{n}$ are lightcone vectors satisfying $n^2 =\overline{n}^2 =0$, $n\cdot \overline{n}=2$.
$Y_n$ and $\tilde{Y}_{\bar{n}}$ are the soft Wilson lines~\cite{Bauer:2001yt,Chay:2004zn} given as
\begin{equation}
Y_n =\sum_{\mathrm{perm}}\exp \Bigl[-g\frac{n\cdot A_s}{n\cdot \mathcal{P} +i0} \Bigr], \  
\tilde{Y}_{\bar{n}} =\sum_{\mathrm{perm}} \exp \Bigl[-g\frac{\overline{n}\cdot A_s}{\overline{n}\cdot \mathcal{P}-i0}\Bigr],
\end{equation}
where $A_s$ is the soft gauge field, and $\mathcal{P}$ is the operator extracting the momentum from the gauge field.
The one-loop corrections for the soft parts are given, using the dimensional regularization for 
both UV and IR divergences, as
\begin{eqnarray}\label{soft1}
S_{\mathrm{DY}}^{(1)}  &=& 
\frac{\alpha_s C_F}{\pi} \Bigl\{ \delta (1-z) \Bigl[ -\frac{1}{\euv^2} +\frac{2}{\euv \eir}+\frac{1}{\eir} \ln \frac{\mu^2}{Q^2} \nonumber \\
&+& \frac{1}{2} \ln^2 \frac{\mu^2}{Q^2} -\frac{\pi^2}{4}\Bigr] -2 \Bigl( \frac{1}{\eir}+\ln \frac{\mu^2}{Q^2}\Bigr) \frac{1}{(1-z)_+} \\
&+&4 \Bigl( \frac{\ln (1-z)}{1-z}\Bigr)_+\Bigr\}, \nonumber  \\
S_{\mathrm{DIS}}^{(1)}  &=& \frac{\alpha_s C_F}{\pi} \Bigl(\frac{1}{\euv} -\frac{1}{\eir}\Bigr) \Bigl[ \delta (1-z) \Bigl(-\frac{1}{\euv}
-\ln \frac{\mu}{Q}\Bigr) \nonumber \\
&+&\frac{1}{(1-z)_+}\Bigr].
\end{eqnarray}
As can be seen clearly, the soft functions involve both UV and IR divergences, therefore the evolution of the soft function at this stage does not
make sense. Furthermore, the term $1/(1-z)_+$ definitely comes from the real gluon emission. In DY process, it is IR divergent
contrary to the belief that it originates from UV divergence.

In employing SCET, it is critical to separate the degrees of freedom into collinear and soft parts carefully. Special care should be taken in 
considering loop corrections in the collinear part since the loop momentum can reach the soft region which should be removed from the
collinear part. In SCET, the zero-bin subtraction \cite{Manohar:2006nz} 
is devised to remove the soft modes from the collinear part 
to avoid double counting. The idea is to take each
Feynman diagram for the collinear contributions, and subtract the contribution when the loop momentum becomes soft. Only after
the zero-bin subtraction, the separation of the collinear and soft parts becomes complete.  We emphasize that the zero-bin contribution 
should have commensurate scales with the  soft part. This needs some elaboration. 
There is a hierarchy of scales $Q \gg E \gg \Lambda_{\mathrm{QCD}}$. We first 
construct $\mathrm{SCET}_{\mathrm{I}}$ between $Q$ and $E$ and the soft zero-bin contribution is that of the collinear part with momentum
scaling as $Q\lambda \sim E$. There might be the ultrasoft (usoft) contribution of order  $Q\lambda^2$ to be subtracted, but it is not
the right choice.  If we naively subtract the usoft modes
from the collinear part, there is mismatch between the soft part in Eqs.~(\ref{sdy}), (\ref{sdis}) and the zero-bin subtraction. In $\mathrm{SCET}_{\mathrm{II}}$ below
$E$, we match each quantity at the boundary $E$, and evolve down to some renormalization scale $\mu$.  Here 
in $\mathrm{SCET}_{\mathrm{II}}$, we also perform loop corrections, but the zero-bin contribution 
here corresponds to the usoft contribution of order $Q\lambda^2$. 

The initial collinear function $f$ in $\mathrm{SCET}_{\mathrm{I}}$ is defined  as
\begin{equation} \label{pdf1}
f(x,\mu) = \langle N | \overline{\chi}_n \frac{\FMSlash{\overline{n}}}{2} \delta (Qx  -
\overline{n}\cdot \mathcal{P}) \chi_n|N\rangle,
\end{equation}
where $\chi_n = W_n^{\dagger} \xi_n$ is the collinear gauge-invariant fermion with the collinear Wilson line $W_n$ given by
\begin{equation} \label{cowil}
W_n =\sum_{\mathrm{perm}} \exp \Bigl[-g\frac{\overline{n}\cdot A_n}{\overline{n}\cdot \mathcal{P}+\delta +i0} \Bigr].
\end{equation}•
The $\delta$ in Eq.~(\ref{cowil}) is the regulator to treat rapidity divergence, if there is any.
It should be matched to the PDF $\phi$ in $\mathrm{SCET}_{\mathrm{II}}$ as
\begin{equation} \label{mpdf}
f(x,\mu) =\int_x^1 \frac{dz}{z} K(z,\mu) \phi \Bigl(\frac{x}{z},\mu\Bigr).
\end{equation}
The PDF $\phi$ at the operator level is the same as Eq.~(\ref{pdf1}) for $f$, but at the scale $\mu \ll E$.
The matching coefficient $K(z,\mu)$ can be computed
by comparing different kinds of zero-bin subtractions  in $\mathrm{SCET}_{\mathrm{I}}$ and $\mathrm{SCET}_{\mathrm{II}}$.

Now we explain the distinction between the soft and usoft zero-bin contributions. In Eqs.~(\ref{sdy}) and (\ref{sdis}), the soft functions
include delta functions. The soft contribution means that $i\partial_0/Q$ or $i\overline{n}\cdot \partial /Q$ is of order $1-z$, and the delta
functions play a nontrivial role in loop computations. If $i\partial_0/Q$ or $i\overline{n}\cdot \partial /Q$ is much smaller than $1-z$, it
can be neglected, and the delta function becomes $\delta (1-z)$. The remaining soft Wilson lines cancel to yield 1. The usoft contribution
corresponds exactly to this case. 

The distinction also applies to the collinear function $f$ and the PDF $\phi$. In Eq.~(\ref{pdf1}), there also 
appears a delta function. Now consider the zero-bin contribution, in which the loop momenta become small compared to the label momenta.
Near the endpoint, the leading momentum almost cancels. Therefore  in real gluon
emissions, the small subleading momentum of order $Q\lambda$ should be kept in the delta function. The soft zero-bin contribution is the
soft limit of the collinear loop calculations of order $Q\lambda$. For $\phi$, we take the limit 
where the loop momentum 
approaches $Q\lambda^2$, and the usoft momentum is neglected in the delta function.  
Therefore the usoft zero-bin contributions are always proportional to $\delta (1-z)$.  

At one loop, $f^{(1)}$  at the parton level is given by
\begin{eqnarray}
f^{(1)}(x,\mu) & =&\tilde{f}^{(1)} - f_{\mathrm{s},0}^{(1)} \\
&= &\frac{\alpha_s C_F}{2\pi} \Bigl(\frac{1}{\euv} -\frac{1}{\eir}\Bigr)  \nonumber \\
&\times& \Bigl[ \Bigl( \frac{2}{\euv}  +2\ln \frac{\mu}{Q}+\frac{3}{2}\Bigr) \delta (1-x) -1-x\Bigr], \nonumber
\end{eqnarray}
where $\tilde{f}^{(1)} $ is the naive one-loop result for Eq.~(\ref{pdf1}), and $f_{\mathrm{s},0}^{(1)}$ is the soft zero-bin contribution.
The PDF $\phi^{(1)}$ at the parton level is given as
\begin{eqnarray} \label{phi1}
\phi^{(1)} (x,\mu)&=& \tilde{f}^{(1)} - f_{\mathrm{us},0}^{(1)} \\
&=&\frac{\alpha_s C_F}{2\pi} \Bigl(\frac{1}{\euv} -\frac{1}{\eir}\Bigr) \Bigl[ \frac{3}{2}\delta (1-x) +\frac{1+x^2}{(1-x)_+}\Bigr],
\nonumber
\end{eqnarray}
where $ f_{\mathrm{us},0}^{(1)}$ is the usoft zero-bin contribution.
Note that $f_{\mathrm{us},0}^{(1)}=0$ and $f_{\mathrm{s},0}^{(1)} = \phi^{(1)} -f^{(1)}$. 
The PDF $\phi$ obeys the standard Dokshitzer-Gribov-Lipatov-Altarelli-Parisi (DGLAP)  evolution equation from the UV behavior of Eq.~(\ref{phi1}).
Near the endpoint for DIS and DY processes, the initial-state jet function $K^{(1)}$  at one loop 
is given by 
\begin{eqnarray}
K^{(1)} (E,\mu) &=& f^{(1)} -\phi^{(1)} =f_{\mathrm{us},0}^{(1)} -f_{\mathrm{s},0}^{(1)}\\
&=&\frac{\alpha_s C_F}{\pi} \Bigl(\frac{1}{\euv} -\frac{1}{\eir}\Bigr) \nonumber \\
&\times& \Bigl[ \delta (1-z) \Bigl( \frac{1}{\euv} 
+\ln \frac{\mu}{Q} \Bigr) -\frac{1}{(1-z)_+}\Bigr]. \nonumber
\end{eqnarray}
Since the collinear contributions for $f$ and $\phi$ are the same, $K$ 
actually comes from the difference between the soft and the usoft zero-bin contributions. Furthermore, since the usoft zero-bin
contributions vanish, $K$ comes from the soft zero-bin subtraction. Note that $K^{(1)}$ also includes the IR divergence.

The final-state jet function $J$ is IR finite, and contains only the UV divergence. Including the zero-bin subtraction it has been computed in~\cite{Idilbi:2007ff}.
The one-loop results show explicitly that the collinear and soft parts except the final-state jet function include IR divergences, 
which hinders the factorization theorem. Therefore
the operator definitions of the soft and collinear parts are not enough to guarantee the factorization. 

We propose to combine the initial-state jet functions and the soft function to obtain an IR finite kernel. 
In $\mathrm{SCET}_{\mathrm{II}}$ below $E$, the scattering cross sections  in Eq.~(\ref{nfac}) can be expressed as follows:
For DY process, it is schematically written as
\begin{equation} \label{facdy}
\begin{split}
d\sigma_{\mathrm{DY}} &=H(Q,\mu) \otimes S(E,\mu) \otimes K_1 (E,\mu) \\
&\otimes K_2 (E,\mu) \otimes 
\phi_1 (\mu)\otimes \phi_2 (\mu) \\
&= H(Q,\mu) \otimes    W_{\mathrm{DY}} (E,\mu) \otimes \phi_1 (\mu)\otimes \phi_2 (\mu),
\end{split}
\end{equation}
where the kernel $W_{\mathrm{DY}}$ is defined as
\begin{equation}
W_{\mathrm{DY}} = S(E,\mu) \otimes K_1 (E,\mu) \otimes K_2 (E,\mu).
\end{equation}
For DIS, it is written as
\begin{eqnarray} \label{facdis}
d\sigma_{\mathrm{DIS}} &=&H(Q,\mu) \otimes J (Q\sqrt{1-z},\mu) \otimes S(E,\mu)  \nonumber \\
&\otimes& K (E,\mu) \otimes \phi (\mu) \\
&=& H(Q,\mu) \otimes J (Q\sqrt{1-z},\mu)\otimes    W_{\mathrm{DIS}} (E,\mu) \otimes \phi (\mu), \nonumber 
\end{eqnarray}
where the kernel $W_{\mathrm{DIS}}$ is defined as
\begin{equation}
W_{\mathrm{DIS}} =S(E,\mu) \otimes K (E,\mu) .
\end{equation}
And the hard factor is given by $H(Q,\mu)=|C(Q,\mu)|^2$, where $C(Q,\mu)$ is the Wilson coefficient of the back-to-back current in SCET.

These kernels are free from IR divergences, and can be computed order by order in perturbation theory. From the viewpoint of SCET,
the kernels are the matching coefficients between $\mathrm{SCET}_{\mathrm{I}}$ and $\mathrm{SCET}_{\mathrm{II}}$,  
free of IR divergences, and should depend only on $Q$ and $E$, or $1-z$.
And the renormalization group behavior
can be probed for the hard part, and the kernel without employing complicated methods to handle IR divergences.
To one loop, the kernels are given as
\begin{equation}
\begin{split}
W_{\mathrm{DY}} &= \delta (1-z) \\
&+\frac{\alpha_s C_F}{\pi} \Bigl\{ \delta (1-z) \Bigl[ \frac{1}{\euv^2} +\frac{1}{\euv} \ln \frac{\mu^2}{Q^2}\\
&+ \frac{1}{2}\ln^2 \frac{\mu^2}{Q^2} -\frac{\pi^2}{4}\Bigr]+4\Bigl(\frac{\ln (1-z)}{1-z}\Bigr)_+\\
&-\frac{2}{(1-z)_+} \Bigl( \frac{1}{\euv} +\ln \frac{\mu^2}{Q^2}\Bigr) \Bigr\}, \\
W_{\mathrm{DIS}} &=\delta (1-z).
\end{split}
\end{equation}

Eqs.~(\ref{facdy}) and (\ref{facdis}) are our factorization formulae for the DY and DIS processes near the endpoint.
As in the conventional factorization theorems, we first start with the factorization of the hard, collinear and 
soft parts. The collinear and soft parts are defined in terms of the matrix elements of gauge-invariant operators. 
Due to the presence of the IR divergences in each part, the collinear and soft parts are reorganized in such a way
that the kernel $W$ and the collinear part $J$ in DIS are infrared finite. In this procedure, the matching 
coefficient of  $f$ and the PDF between
$\mathrm{SCET}_{\mathrm{I}}$ and $\mathrm{SCET}_{\mathrm{II}}$ is included in the kernel $W$. In the form of 
Eqs.~(\ref{facdy}) and (\ref{facdis}), the kernel and the hard part are physically meaningful and the evolution of these
can be considered unambiguously.

If a process involves small transverse momentum, the relevant  collinear function satisfies
 \cite{Chay:2012mh}
\begin{equation}
f(x,\mathbf{k}_{\perp},\mu) = \int_x^1 \frac{dz}{z} K_{\perp} (z,\mathbf{k}_{\perp}, \mu) 
\phi \Bigl(\frac{x}{z},\mu\Bigr).
\end{equation}
Combining the transverse-momentum-dependent soft function with $K_{\perp}$, 
we can also obtain an IR finite kernel $W_{\perp}$ 
as $W_{\mathrm{DY}}$ is obtained near the endpoint. In this case, there appears
an additional divergence called the rapidity divergence. The rapidity divergence can be regulated by a rapidity regulator 
\cite{Chiu:2009yx}, and it is shown that the rapidity divergence cancels in each collinear sector if the regulator is introduced only 
in the collinear Wilson lines \cite{Chay:2012mh}. It is also true near the endpoint.

Remarkably the kernel for DIS becomes $\delta (1-z)$ to one loop since $S^{(1)}_{\mathrm{DIS}} + K^{(1)}=0$, 
and the inclusive DIS cross section near the endpoint consists of the hard part, the jet 
function and the PDF $\phi$ at the scale $\mu <E$. It means that the soft contribution
and the initial-state jet function cancel. Therefore
the PDF satisfies the ordinary evolution equation, as in full QCD. If we separate the soft contribution and the collinear 
contribution, the corresponding 
collinear part  $f=K\otimes \phi$ does not satisfy the DGLAP equation, which was discussed in previous literature~\cite{Chay:2005rz,Idilbi:2007ff,Fleming:2012kb}.
The fact that $W_{\mathrm{DIS}}=\delta (1-z)$ is shown to be true to all orders in $\alpha_s$ 
by a simple argument. The usoft zero-bin contributions vanish in $\phi$. The soft zero-bin contribution from $f$
is obtained by integrating out the momenta of order $Q \lambda$. This is exactly the procedure to obtain the eikonal
form of the soft Wilson line in the $\overline{n}$ direction. Therefore the initial-state jet function, which is the negative value of the 
soft zero-bin contributions, always cancels the soft function to all orders in $\alpha_s$. 
In contrast, it is different in DY processes since the soft function involves
an interaction between $n$ and $\overline{n}$-collinear fermions, while the collinear part which interacts only within each collinear sector
does not produce the same soft interaction in the zero-bin limit. 

Now that the explicit results of factorization are presented, the explanation about the essential idea of our factorization 
formula is in order. The basic idea of
decoupling the collinear and soft parts in the conventional factorization scheme is still respected here, 
but an important step should be added to complete
factorization. The separation of the 
collinear and soft parts is implicit in the conventional factorization scheme, however it is necessary to remove the soft limit 
 in the collinear part properly at higher orders.
 It appears in loop calculations of the collinear part, and the contribution from the loop momentum in the soft (or usoft) limit 
should be removed since it is already included in the soft part. Our procedure of factorization exactly follows this idea. The soft 
contribution in the collinear part
in loop calculations is extracted by the zero-bin subtraction, and there is no double counting of the soft degrees of freedom.

As a result, 
the mixing of IR and UV divergences disappears, and the IR divergences are transformed to the UV divergences through the zero-bin subtraction
as long as the zero-bin subtractions are performed properly.  And the result for the kernel turns out to be IR finite. In this line of thought,
the jet function in DIS takes a special place since it becomes IR finite after the zero-bin subtraction. But considering the fact that the jet 
function is the matching coefficient, it is obvious that the jet function is IR finite.

Let us summarize how the factorization theorems proceed in our scheme. First, construct 
the scattering amplitude, or the relevant matrix elements at the operator level to separate the hard,
collinear and soft parts. Second,  extract the soft or usoft contributions
depending on the appropriate energy scales by the zero-bin subtraction. Third, combine the
contributions from the soft function and the initial-state jet function from the
difference of the zero-bin subtractions in the collinear part. The combined sum constitutes the IR-finite 
kernel $W$. As a result, the hard part and the kernel $W$   can be computed
in perturbation theory, and the evolution can be set up without concerns on how to treat IR divergences in each part. 
The detailed calculation will be presented in Ref.~\cite{Chay:2013zya}.   
This factorization method has been also applied to the processes with small transverse momentum \cite{Chay:2012mh}.  It strongly implies
that our factorization idea can be widely used in various high-energy scattering processes.

J.~Chay  and C.~Kim were supported by Basic Science Research Program through the National Research Foundation of Korea (NRF)
funded by the Ministry of Education, Science and Technology (No. 2012R1A1A2008983 and No. 2012R1A1A1003015),  respectively.

\end{document}